\documentclass[prl,aps,twocolumn,showpacs,floatfix]{revtex4}
\usepackage{epsfig}
\usepackage{graphicx}
\usepackage[dvips]{color}

\newcommand{\beq}{\begin{displaymath}}
\newcommand{\eeq}{\end{displaymath}}
\newcommand{\ben}{\begin{equation}}
\newcommand{\een}{\end{equation}}
\newcommand{\bea}{\begin{eqnarray}}
\newcommand{\eea}{\end{eqnarray}}
\def\bra{\langle }
\def\ket{\rangle}

\begin{document}
\author{A. Fiasconaro$^1$\footnote{E-mail address: afiasconaro@gip.dft.unipa.it},
 B. Spagnolo$^1$ and S. Boccaletti$^2$}
\affiliation{$^1$ Dipartimento di Fisica e Tecnologie Relative and
INFM-CNR, Group of Interdisciplinary Physics\footnote
{http://gip.dft.unipa.it},
\\ Universit\`a di Palermo, Viale delle Scienze pad.~18, I-90128 Palermo, Italy \\
$^2$ Istituto  dei Sistemi Complessi del CNR, Sezione di Firenze,
\\ Via Madonna del Piano, Sesto Fiorentino, Florence, Italy}
\title{Signatures of Noise Enhanced Stability in Metastable States}

\begin{abstract}
The lifetime of a metastable state in the transient dynamics of an
overdamped Brownian particle is analyzed, both in terms of the
mean first passage time and by means of the mean growth rate
coefficient. Both quantities feature non monotonic behaviors as a
function of the noise intensity, and are independent signatures of
the noise enhanced stability effect. They can therefore be
alternatively used to evaluate and estimate the presence of this
phenomenon, which characterizes metastability in nonlinear
physical systems.
\end{abstract}
%
%

\maketitle

Metastability is a generic feature of many nonlinear systems, and
the problem of the lifetime of a metastable state involves
fundamental aspects of non-equilibrium statistical mechanics.
Metastable states, indeed, have been proved to play a crucial role
e.g. in protein folding dynamics, Ising spin glasses, complex
dynamics of large molecules at surfaces, enhancement of cellular
memory, and in dynamics of cellular reactive oxygen
species~\cite{Dob}. The problem of the lifetime of a metastable
state has been addressed in a variety of areas, including
first-order phase transitions, Josephson junctions, field theory
and chemical kinetics~\cite{Cha}.

Recently, the investigation of nonlinear dynamics in the presence
of external noisy sources led to the discovery of some
resonance-like phenomena, among which  we recall stochastic
resonance~\cite{Gam}, resonant activation~\cite{Doe}, and noise
enhanced stability (NES)~\cite{Man,Hir,Agu}. All these phenomena
are characterized by a nonmonotonic behavior of some quantity as a
function of the forcing noise intensity or the driving frequency,
that reflects a constructive and counterintuitive effect of the
noise acting on the nonlinear system. In particular, several
theoretical studies have shown that the average escape time from a
metastable state in fluctuating and static potentials has a non
monotonic behavior as a function of the noise
intensity~\cite{Agu,Ale,Mal}. This resonance-like behavior, which
contradicts the monotonic behavior predicted by the Kramers theory
\cite{Kra,Han}, is called the NES phenomenon: the stability of
metastable or unstable states can be enhanced by the noise and the
average life time of the metastable state is larger than the
deterministic decay time. Furthermore, if more realistic noise
sources (such as colored noise with a finite correlation time) are
considered, the value of the noise intensity at which the maximum
of the average escape time occurs is even larger than that
corresponding to the white noise case, meaning that the NES effect
could be easy to measure experimentally, because of the finite
time correlations involved in any realistic noisy
sources~\cite{Ale}.

When considering a Brownian particle in the presence of a
metastable potential, the NES  effect is always obtained
\cite{Man}, regardless on the unstable initial position of the
particle. More precisely, two different dynamical regimes occur:
one is characterized by a non monotonic behavior of the average
escape time, as a function of noise intensity, and the other
features a divergence of the mean escape time when the noise
intensity tends to zero, implying that the Brownian particle
remains trapped within the metastable state in the limit of small
noise intensities. The description of the transition from one
dynamical regime to the other is yet an open question.

\begin{figure}[htbp]
 \begin{center}
  \includegraphics[height=10cm]{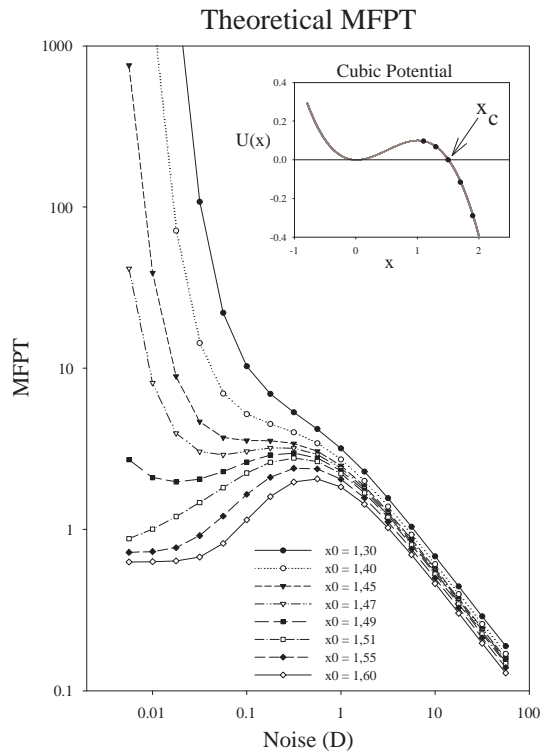}
\caption{MFPT  $\tau(x_0,x_F)$ evaluated from expression
(\ref{double int}) {\it vs.} the noise intensity $D$, for
$x_F=2.2$ and for different initial positions $x_0$ (see the
legend). The NES effect is observed for $x_0 > x_c$, while the
divergence regime corresponds to for $x_{u} < x_0 < x_c$. Inset:
the cubic potential $U(x)$ with the metastable state at $x=0$. The
arrow indicates the intersection point $x_c=1.5$ between the
potential curve and the horizontal axis.}
  \label{th}
 \end{center}
\end{figure}

In this Letter we analyze in more detail the divergent dynamical
regime, and suggest an approach for detecting the stability of
metastable states, that is alternative to the
mean-first-passage-time (MFPT) technique. In particular, we find a
non monotonic behavior of the MFPT with a minimum and a maximum as
a function of the noise intensity for initial positions of the
Brownian particle close to the point $x_c$ where the potential
shape intersects the $x-$axis (see the inset of Fig.~\ref{th}). In
this regime the standard deviation of the escape time has a
divergent behavior for small noise intensity. This means that the
average escape time, that is a quantitative measure of the average
life time of the metastable state, has in fact some statistical
limitations to fully describe the stability of metastable states.

To complement the analysis of the transient dynamics of metastable
states, we then introduce a different approach, based on the
evaluation of the mean growth rate coefficient $\Lambda$ as a
function of the noise intensity. The $\Lambda$ coefficient is
evaluated by the use of a procedure similar to that for the
calculation of the Lyapunov exponent in stochastic systems, e.g.
we consider the evolution of the separation $\delta x(t)$ between
two neighboring trajectories of the Brownian
particle~\cite{Ebe,Pal,Wit}. We will show that also the $\Lambda$
coefficient displays a non monotonic behavior (with a clear
minimum), as a function of the noise intensity, thus representing
an independent way for detecting and estimating the NES effect.

The starting point of our analysis is a Brownian particle in one
spatial dimension, obeying the following Langevin equation

 \ben
 \label{eq}
  \dot{x}=-\frac{d U(x)}{d x} + \sqrt{D} \xi(t),
 \een
 where $\xi(t)$ is a white Gaussian noise source with zero mean and
 $\delta$-correlated in time
 ( $\bra \xi(t) \ket = 0$ and $\bra \xi(t)\xi(t+\tau) \ket = \delta
 (\tau)$), and
$  U(x)=0.3x^2 - 0.2x^3$ is a cubic potential, whose shape is
shown in the inset of Fig.~\ref{th}.  The potential profile has a
local stable state  at $x_s=0$, an unstable state at $x_u=1$, and
intersects the $x$-axis at $x_c=1.5$.

After fixing a given target position $x_F > x_c$, the MFPT
$\tau(x_0,x_F)$ (the average time for a particle starting from an
initial position $x_0$ to reach $x_F$) is given by the closed
analytical expression~\cite{Han,Gar}

\vskip -0.2cm
 \ben
 \label{tau}
  \tau(x_0,x_F)=\frac{2}{D} \int_{x_0}^{x_F} e^{2u(z)}
  \int_{-\infty}^{z} e^{-2u(y)} \;\; dy dz \ ,
 \een
 where $u(x)$ is the dimensionless potential profile,
 obtained by normalizing $U(x)$ to the noise intensity $D$
 ($u(x)=\frac{U(x)}{D}$).
The double integral (\ref{tau}) can be evaluated (in part
analytically, in part numerically), giving rise to

\ben \label {double int}
 \tau(x_0,x_F) = \frac{2}{D}\int_{x_0}^{x_F} e^{2u(z)}
G(z) dz,
 \een
where $G(z) = 0.6046~e^{-z} [I_{-1/3}(z)+I_{1/3}(z)]
 -\frac{1}{2}  \ {}_{2}F_{2}({\frac{1}{2},1};{\frac{2}{3},\frac{4}{3}};-2z)
 + \int_{0}^{z} e^{-2u(y)}\; dy$, $z = 1/(10D)$, $I_n(z)$
 is the modified Bessel function of the first kind and
 $_{p}F_{q}(a_1,a_2;b_1,b_2;z)$ is the generalized
hypergeometric function.

In Fig.~\ref{th} the evaluation of expression (\ref{double int})
for $x_F=2.2$ and for different initial positions $x_0$ (as
sketched in the legend) is reported. Two different regimes can be
observed, depending on the initial state of the particle: {\it
(i)} the NES effect for all initial conditions $x_0 > x_c$, and
{\it (ii)} the divergent regime for $x_u \leq x_0 < x_c$, where
$x_u=1$ is the location of the relative maximum of the potential
profile.

In this latter regime, for all initial positions $x_0$ smaller
than (but sufficiently close to) $x_c$, the MFPT displays a non
monotonic behavior with a minimum and a maximum. For very low
noise intensities, the Brownian particle is trapped into the
potential well, as a consequence of the divergence of the MFPT in
the limit $D \rightarrow 0$. For increasing noise intensity, the
particle can escape out more easily, and the MFPT decreases. As
the noise intensity reaches a value $D \approx \Delta U = 0.1$
(corresponding to the potential barrier height), the concavity of
the MFPT curves changes. Close to such a noise intensity, the
escape process of the Brownian particle is slowed down, due to the
fact that the probability to reenter the well is increased. At
higher noise intensities one recovers a monotonic decreasing
behavior of the MFPT. In summary, from inspection of
Fig.~\ref{th}, the behavior of MFPT {\it vs.} $D$ goes with
continuity from a monotonic divergent behavior to a non monotonic
finite behavior (typical NES effect), passing through a non
monotonic divergent behavior with a minimum and a maximum.

\begin{figure}[htbp]
 \begin{center}
  \includegraphics[height=6cm]{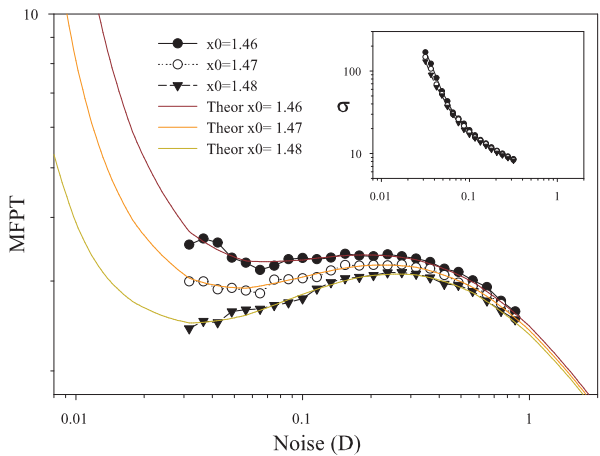}
\caption{Numerical estimation of the MFPT (symbols) {\it vs.} $D$
for $x_0 = 1.46$ (filled circles), $x_0= 1.47$ (empty circles),
and $x_0=1.48$ (filled triangles). The solid curves are the
corresponding estimations of the MFPT from Eq.~(\ref{double int}).
Inset: standard deviation $\sigma$ of MFPT's corresponding to
different noise realizations {\it vs. } $D$ for the same initial
positions $x_0$.}
  \label{sim}
 \end{center}
\end{figure}

In the following we will focus on this last dynamical regime.
First of all, we compare the theoretical results of Fig. \ref{th}
with direct numerical simulations of Eq. (\ref{eq}). Namely, we
numerically integrate Eq. (\ref{eq}) with different initial
conditions $x_0 \lesssim x_c$. For each initial condition (and for
each value of the noise intensity $D$) , the integration is
performed over an ensemble of $N_R=350,000$ different realizations
of the white noise process $\xi(t)$. The target position $x_F=2.2$
is selected, and the MFPT (calculated as the ensemble average of
the first passage times through $x_F$ for different noise
realizations) is reported in Fig.~\ref{sim}, together with the
corresponding theoretical curves. The agreement between
theoretical predictions and numerical simulations is very good.
For low values of $D$, however, we cannot reproduce the
theoretical curves because of the finiteness of the integration
time ($T_{max} = 20,000~a. u.$) and of the number of realization.
The MFPT evaluated in this range of noise intensities, therefore,
tends to the deterministic escape time.

The numerical study of Eq. (\ref{eq}) allows also to calculate the
standard deviation $\sigma$ of the set of first passage times
obtained for different noise realizations. The results are shown
in the inset of Fig.~\ref{sim} for the same values of $x_0$, where
a divergent behavior of $\sigma (D)$ is visible in the limit $D
\rightarrow 0$. Such a feature confirms that the only information
on the MFPT is not sufficient to fully unravel the statistical
properties of this dynamical regime, and motivated our search for
a complementary approach.

This is done by monitoring the properties of the {\it mean growth
rate} coefficient $\Lambda$. Let $\delta x_0= \delta x (t=0) \ll
1$ be the initial separation of two neighboring Brownian particles
subjected to the same noise process $\xi(t)$. By linearization of
Eq. (\ref{eq}), the evolution of the particle separation  $\delta
x(t)$ is given by

 \ben
 \label{delta}
  \dot{\delta x(t)}=-\frac{d^2U(x)}{dx^2} \delta x(t)=
  \lambda_i(x,t) \delta x(t) \ ,
 \een
and allows for the definition of an  instantaneous growth rate
$\lambda_i(x,t)$. It is important to stress that, in Eq.
(\ref{delta}), $\frac{d^2U(x)}{dx^2}$ is calculated onto (and as
so, it is a function of) the noisy trajectory $x[\xi(t)]$
\cite{comment}. The growth rate coefficient $\Lambda_i$ (for the
$i_{th}$ noise realization), is then defined as the long-time
average of the instantaneous $\lambda_i$ coefficient over
$\tau(x_0,x_F)$ \cite{Wit}
 \ben
 \label{lyap}
  \Lambda_i = \frac{1}{\tau(x_0,x_F)} \int_{0}^{\tau(x_0,x_F)} \lambda_i(x,s)
ds.
 \een

Notice that, in the limit $\tau(x_0,x_F) \to \infty$, Eq.
(\ref{lyap}) coincides formally with the definition of the maximum
Lyapunov exponent, and that, therefore, our $\Lambda_i$
coefficient has the meaning of a finite-time Lyapunov exponent,
since we are interesting in characterizing a transient dynamics.
The mean growth rate coefficient (MGRC) $\Lambda$ is then obtained
by ensemble averaging the $\Lambda_i$ coefficients over the $N_R$
different noise realizations ($ \Lambda=\frac{\sum
\Lambda_i}{N_R}$).

In our simulations of Eq. (\ref{delta}), an integration time step
of $dt = 0.0001$, and an initial condition $\delta x_0=0.01$ are
used. The instantaneous growth rate $\lambda_i(x,t)$ is calculated
over $N_{\tau}$ small subintervals of the trajectory, each of them
corresponding to a time interval $\tau = 5 dt$.

\begin{figure}[htbp]
 \begin{center}
  \includegraphics[height=4cm]{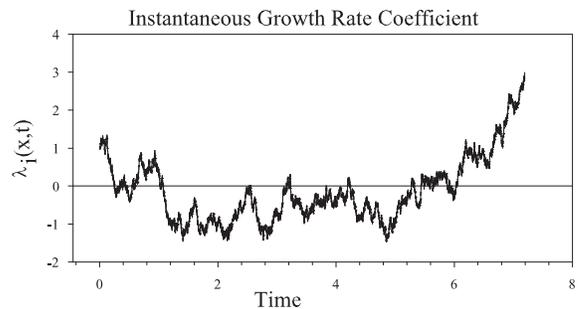}
 \caption{$\lambda_i(x,t)$ (see text for definition) {\it vs.} time, for
 $x_0 = 1.3$, $x_F = 3.0$, and $D = 1$.}
 \vskip -0.3 cm
  \label{lambdat}
 \end{center}
\end{figure}

\begin{figure}[htbp]
 \begin{center}
  \includegraphics[height=10cm]{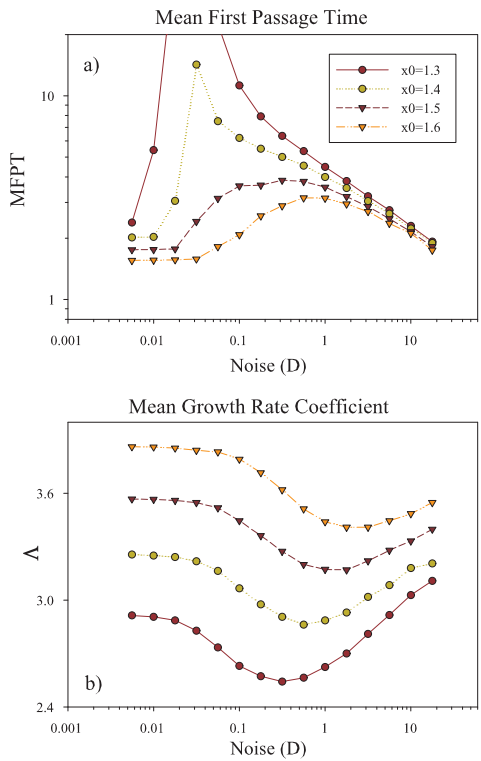}
\caption{(a) Numerical estimation of the MFPT for various initial
positions (as in the legend) and (b) the corresponding MGRC {\it
vs.} the noise intensity $D$. The average has been performed over
$20,000$ realizations. The absorbing boundary is $x_F=20.0$, and
the maximum waiting time used in the simulations is
$T_{max}=20,000$.}

  \label{lya}
 \end{center}
\end{figure}

In Fig.~\ref{lambdat} we report the behavior of  $\lambda_i(x,t)$
as a function of time for $D = 1$. The $\lambda_i(x,t)$
coefficient is negative for most of the time, due to noise-induced
local stability of metastable state. As the Brownian particle
reaches the point at which the potential changes its concavity,
$\lambda_i(x,t)$ becomes positive. As for the $\Lambda$
coefficient, fixing $x_F=20.0$, it displays a non monotonic
behavior with a minimum as a function of noise intensity (see
Fig.~\ref{lya}b), though being always positive. This reflects the
fact that the Brownian particles  always escape in average from
the metastable state, but the non monotonic behavior of the MGRC
marks the presence of the NES effect.  In particular,  the closer
is the initial position of the particles to $x_u$, the smaller is
the MGRC. Notice that $\lambda_i(x,t)$ is proportional to the
trajectory of the Brownian particle because the potential is a
cubic one. Any metastable state, however, can be described through
a local cubic potential even if the real potential has other local
or global stable states.

Fig.~\ref{lya} is a comparative plot of the numerically calculated
MFPT (a) and the corresponding MGRC (b) {\it vs.} $D$ for various
initial positions of the Brownian particles. The MFPT's plot
[Fig.~\ref{lya} (a)] shows the divergent behavior as a maximum,
which is shifted towards lower values of the noise intensity.
Because of the finiteness of the ensemble of particles considered
in our numerical experiments, observation of the divergence in
time is prevented in our simulations, and for very low noise
intensities ($D \rightarrow 0$) the deterministic escape time is
retrieved. As it can be seen in Fig.~\ref{lya} (b), the maximum of
the MFPT curve is reflected by a local minimum in the MGRC shape,
which is however slightly shifted towards lower values of the
noise intensity. This is because, at low noise intensities, not
all the particles coming back into the potential well reach
positions around the metastable state with positive concavity,
that instead will be attained by particles experiencing larger
noise drivings. These latter ones contribute to lower down the
value of $\Lambda$, due to the negative contributions of the
$\lambda_i$'s that corresponds to such positions (see
Fig.~\ref{lambdat}).

In conclusion, we investigated the average escape time from a
metastable state in a cubic potential profile for different
initial unstable positions of the Brownian particles. The two
introduced measures (the behavior of the MFPT and of the MGRC)
furnish suitable tools to detect the lifetime of metastable states
in the presence of external noise, and, as so, they can be
alternatively used in many relevant circumstances, such as
understanding activation processes in complex systems
characterized by energy landscape \cite{Dob}.

 This work was supported by MIUR-FIRB and INFM-CNR.

\end{document}